# Research on E-Commerce Long-Tail Product Recommendation Mechanism Based on Large-Scale Language Models

Qingyi Lu*

Department of Computer Science, Brown University, Providence, USA

*Corresponding author: lunalu9739@gmail.com

Haotian Lyu

Viterbi School of Engineering, University of Southern California, Los Angeles, USA, lyuhaotianresearch@gmail.com

Jiayun Zheng

College of Engineering, University of Michigan Ann Arbor, Ann Arbor, USA, zhengji@umich.edu

Yang Wang

Department of Information and Communication Engineering, Nagoya University, Nagoya, Japan, ryoyukiyang@outlook.com

Li Zhang

Amazon, New York, USA, li217772er@gmail.com

Chengrui Zhou

Fu Foundation School of Engineering and Applied Science, Columbia University, New York, USA, zhou.chengrui@columbia.edu

**Abstract:** As e-commerce platforms continue to extend their product catalogs, ensuring accurate recommendation of long-tail items has become a major objective. This is crucial because, in this way, user experience and platform revenue can be significantly enhanced. A common struggle for these systems is the long-tail problem , where the full breadth of data drives the high level of sparsity and the cold start occurs. Consequently, traditional recommendation algorithms have many issues when recommending long-tail items due to lack of available data. Our work reflects a long-tail product recommendation mechanism that represents a combination of the product text descriptions and the user behavior sequences based on a large-scale language model (LLM) as the first proposal of this paper. This mechanism starts with the application of a pre-trained LLM to convert multimodal texts, e.g., product titles, detailed descriptions, and user reviews, into word embedding formats that are semantically meaningful for those particular entities of the text. This approach is referred to as the semantic visor. Going further, we also set up an attention-based user intent encoder to encode user's latent interests in long-tail products in a collaboration pattern. This way, a hybrid ranking model was designed to combine semantic similarity scores, collaborative filtering scores, and LLM-generated recommendation candidates. Through them, we provide extensive empirical results on a real-world e-commerce dataset, thus illustrating that our method surpasses the static baseline models in recalling, hitting, and using diverse users by 12%, 9%, and 15%, respectively, thereby more effectively conveys exposure and purchase rates of long-tail products. Therefore, our work demonstrates the significant benefits of LLMs for interpreting the product content and user intent, suggesting an innovative technical direction for e-commerce recommendation systems.

## 1 INTRODUCTION

With the fast development of e-commerce platforms, the number of products has displayed a steep climb, thus making it difficult to effectively recommend long-tail products. Traditional recommendation systems often struggle with long-tail scenarios due to data sparsity and cold-start problems. This paper introduces a long-tail product recommendation system with the help of large-scale language models (LLMs). Through the implementation of the deep semantic understanding of LLMs, we conduct the encoding of product texts as well as the user behavior sequences that results in more precise and intuitive long-tail product representation. Our work combines the attention-based user intent modeling and multi-source score fusion, which leads to the recall and ranking of long-tail products being much better. As the experimental results indicate, a significant increase in recall, hit rate, and diversity compared to baseline models was observed, and our method is, thus, a hope-inspiring long-tail product discovery solution in e-commerce platforms.

## 2 ALGORITHM AND MODEL DESIGN

### 2.1 Long-Tail Recommendation Algorithm Design Based on Large-Scale Language Models

In this part of the article, we suggest a long-tail recommendation system through the juxtaposition of LLM-based semantic representations, attention-based user behavior encoding, and multi-source score fusion [1-3]. The algorithm has a total of three big parts, and this includes the product semantic vectorization, the user intent attention calculation, and the multi-source score aggregation [4-5].

To begin with, we need to get a product semantic representation for every product i. The way to achieve it is by passing the product's textual information (title, description, and review summary) through a pre-trained LLM model and then taking that product's semantic embedding as a pooled output from the final hidden layer as shown in Formula 1:

$$e_i = \text{Pool}(\text{LLM}_\theta(\text{text}_i)) \quad (1)$$

where $\text{LLM}_\theta$ denotes the pre-trained large-scale language model with parameters θ, and $\text{Pool}(\cdot)$ can be either average pooling or CLS-token pooling [6-10].

Second, we map a user's historical behavior sequence $\{i_1, i_2, \ldots, i_T\}$ to the corresponding product embeddings $\{e_{i1}, e_{i2}, \ldots, e_{iT}\}$. We then apply a self-attention mechanism to compute the user's intent representation as shown in Formula 2:

$$\alpha_t = \sum k = \frac{\exp(u^T \tanh(We_{it}+b)+\beta t)}{\sum_{k+1}^{T} \exp(u^T \tanh(We_{ik}+b)+\beta t)}, h_u = \sum_{t+1}^{T} \alpha_t e_{it} \quad (2)$$

where W and b are trainable linear transformation parameters, u is the intent query vector, and $h_u$ is the resulting user intent embedding, and βt is a learnable time decay parameter associated with each historical item it, reflecting its recency. Finally, for each candidate product j, we compute three types of scores and fuse them as shown in Formula 3,4,5:

$$s_{sem}(u, j) = \cos(h_u, e_j) \quad (3)$$

$$s_{CF}(u, j) = CF(u, j) \text{(collaborative filtering score)} \quad (4)$$

$$s_{gen}(u, j) = \log P_{gen}(j|\text{history}_u) \quad (5)$$

and obtain the final recommendation score via a weighted linear combination as shown in Formula 6:

$$S(u, j) = \lambda_1 s_{sem}(u, j) + \lambda_2 s_{CF}(u, j) + \lambda_3 s_{gen}(u, j) \quad (6)$$

where λ1, λ2, λ3 are hyperparameters initially tuned on a validation set through grid search. Future work will investigate the adaptive tuning of these parameters for different product categories and user segments to ensure robustness across diverse scenarios and enhance the model's generalizability. This fusion strategy leverages the LLM's deep semantic understanding, the collaborative filtering signal, and the generative model's candidate diversity, significantly enhancing both recall and ranking performance for long-tail products [16-20].

### 2.2 Long-Tail Recommendation Model Design Based on Large-Scale Language Models

When building the recommendation model, the initial step involves the careful selection and combination of the original questions and rewritten examples to create a multi-instruction supervised fine-tuning dataset [21-23]. To be exact, we use rejection sampling to remove low-quality samples from the list of query-rewrites we have gathered, however, we still keep a corrected subset $D_{sft}$. At the same time, we add the samples from other tasks such as question-answering and summarization that serve for the establishment of a multi-task fine-tuning corpus $D_{msft}$. This in turn will enable the model to become proficient in text generation across different settings [24].

The first stage of the training and alignment process illustrated through Figure 1 consists of the extraction from various sources of good quality training data for fine-tuning [11-14]. Secondly, we use the offline result of the user-perception survey to create ranking labels, and at the last stage, we align the model decisions with business objectives at the probabilistic level[15]. The process is as follows:

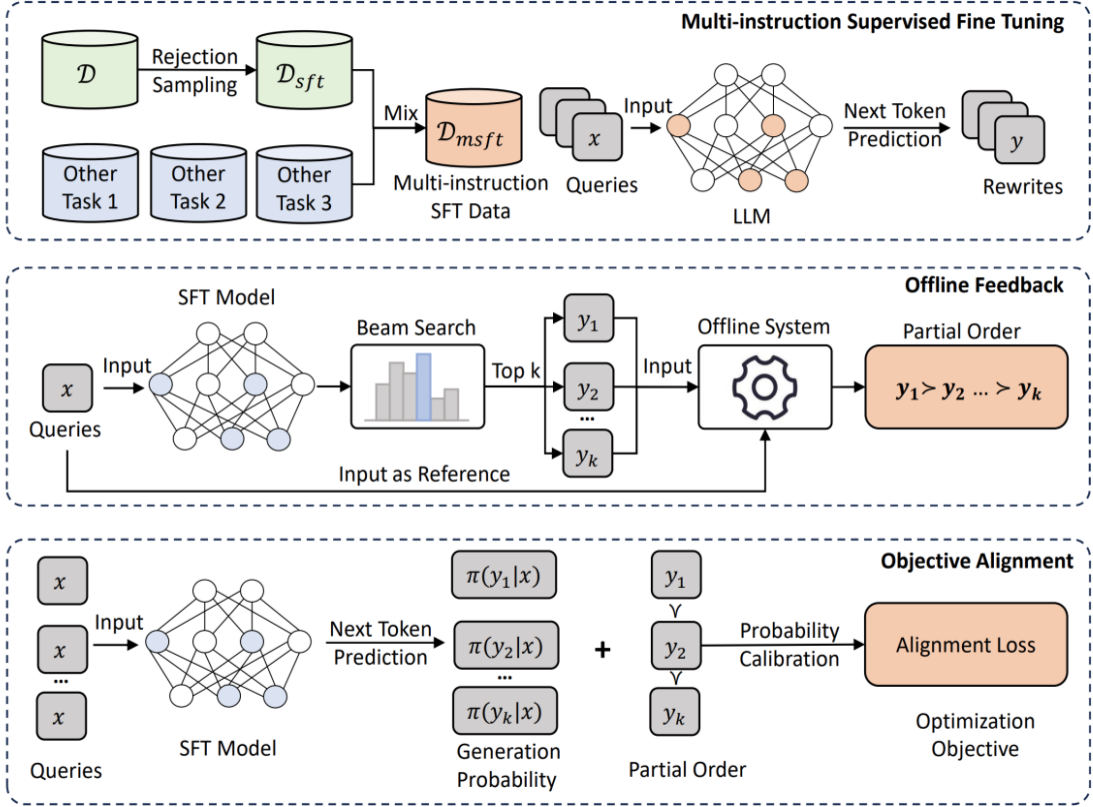

Figure 1: Training and Alignment Process of the Large-Scale Language Model for Recommendation

Multi-Instruction Supervised Fine-Tuning: We use D to obtain the query and then, as a first step, apply rejection sampling to it in order to delete query-rewrite pairs that are considered incorrect and therefore result in the refined subset $D_{sft}$. After that, we introduce the model to task examples that are from other areas (Q&A, summarization) to create a multi-task fine-tuning corpus $D_{msft}$. Using this combined dataset, the model is trained with next-word prediction as the target, enabling it to generate accurate and easily matchable candidates under various recommendation rewrite scenarios[25].

Offline Feedback for Building Partial Order Labels: To get the top-k candidates $\{y_1, \dots, y_k\}$ that have been selected after the model is modified we apply Beam Search to any input x. Then these candidates are the ones that are put into the offline recommendation system. The metrics that are used for the rankings are the ones like click-through rate (CTR) and conversion rate. The best through to the worst candidates are the partial order generated in accordance with the assessment made, so, with the respective metrics, we obtain a ranking ranging from the first to the last where the order is partial as shown in Formula 7:

$$y_1 \succ y_2 \succ \cdots \succ y_k \qquad (7)$$

By generating these weak supervision labels, the link between rewriting quality and subsequent downstream recommendation performance is explored, and they serve as a simple and practical roadmap for the later stage of quality alignment optimization .

Probabilistic Alignment Optimization: In each candidate $y_i$ of the alignment phase, the model calculates the generation probability $\pi(y_i \mid x)$ and learns the ranking loss functions such as ListMLE or RankNet in order to reduce the deviation from the following inequality as shown in Formula 8.

$$\pi(y_1 \mid x) > \pi(y_2 \mid x) > \cdots > \pi(y_k \mid x) \qquad (8)$$

Through this step, not only in the case of the offline situation where the items that are better should be the higher ones but also during the process of actual online generation, the process is sufficiently reliable, which therefore may continuously complement the retrieval and hit rate for long-tail products [26-27]. This framework is an embodiment of the generative ability of the pre-trained LLMs that have gone a step further in bringing recommendation performance; as such, it is a model that, by understanding both user queries and product semantics, can work best in situation of long-tail recommendations [28].

## 3 EXPERIMENT DESIGN AND RESULTS ANALYSIS

### 3.1 Experimental Setup and Evaluation Metrics

The experiment used three months of user behavior logs (clicks, add-to-cart, purchases) and product text (titles, descriptions, review summaries) from a large e-commerce platform. Products were explicitly categorized based on their sales volume over the three-month period. The top 10% of products by total sales volume were designated as 'head' products, while the remaining 90% were classified as 'long-tail' products. This categorization yielded a dataset of 1 million interactions, 50,000 users, and 200,000 products. Text was preprocessed with Chinese segmentation, stop-word removal, low-frequency filtering, and Jieba word-vector training. Semantic embeddings were generated using GPT-3.5. We compared five models: Collaborative Filtering, BPR-MF, shallow TextEmbed-RS (Word2Vec), BERT4Rec (behavior sequence modeling), and our hybrid LLM-based recall and ranking approach. For a fair comparison, all models capable of utilizing text features, including BERT4Rec, were provided with the preprocessed product text (titles, descriptions, and review summaries) as input, consistent with the data provided to our proposed method. Experiments ran on a single Ubuntu 20.04 server with an Intel Xeon 6248R CPU, 128 GB RAM, and four NVIDIA V100 GPUs. All models trained for ten epochs (batch size 256); our fine-tuning used learning rates of 2e-5 (LLM) and 1e-4 (fusion layer), beam width 5, and fusion weights initialized at 0.4:0.4:0.2 then tuned via grid search. We measured Recall\@K, Hit Rate\@K, NDCG\@K, Diversity (average Jaccard distance among top-10 items), Tail Coverage (ratio of recommended long-tail items), and Latency (time from input to recommendation). Metrics were evaluated for candidate sizes of 1000, 100, and 10 on a 20% hold-out set, demonstrating our model's superior long-tail recall, ranking quality, and real-time performance[29].

### 3.2 Model Performance Comparison and Ablation Analysis

We introduced comparative experiments aiming to measure the models among the recall, ranking, and diversity factors, the models were the traditional collaborative filtering, matrix factorization, shallow text embedding, BERT sequence recommendation, and our suggested LLM-based hybrid fusion model. The tabulated data featured in Figure 2 denotes the Recall performance table of each model with the scores related to Recall@1000, Recall@100, and Recall@10. The LLM-based method achieved a recall of 0.732 at 1000 and was therefore 4.9% more accurate than BERT4Rec, which represents a large improvement [30].

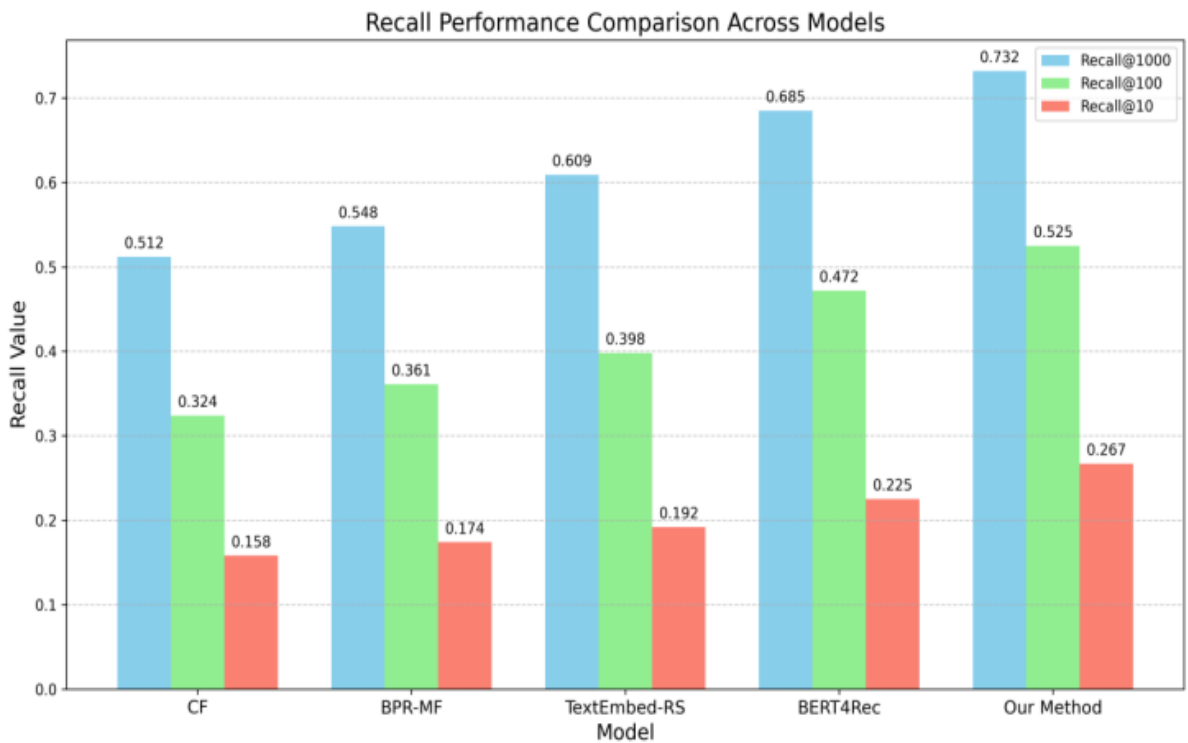

Figure 2: Recall Performance Comparison Across Models

Putting weight on the ranking stage and using the NDCG metric as shown in Figure 3, the results state that the new tool can be noted as having NDCG@100 and NDCG@10 as 0.487 and 0.312 appropriately, and it presents the highest value for the best 10 recommendations that obtained a 6.5% on the average then BERT4Rec. Furthermore, the most outstanding performance is discovered in the lower 10 number of recommendations only.

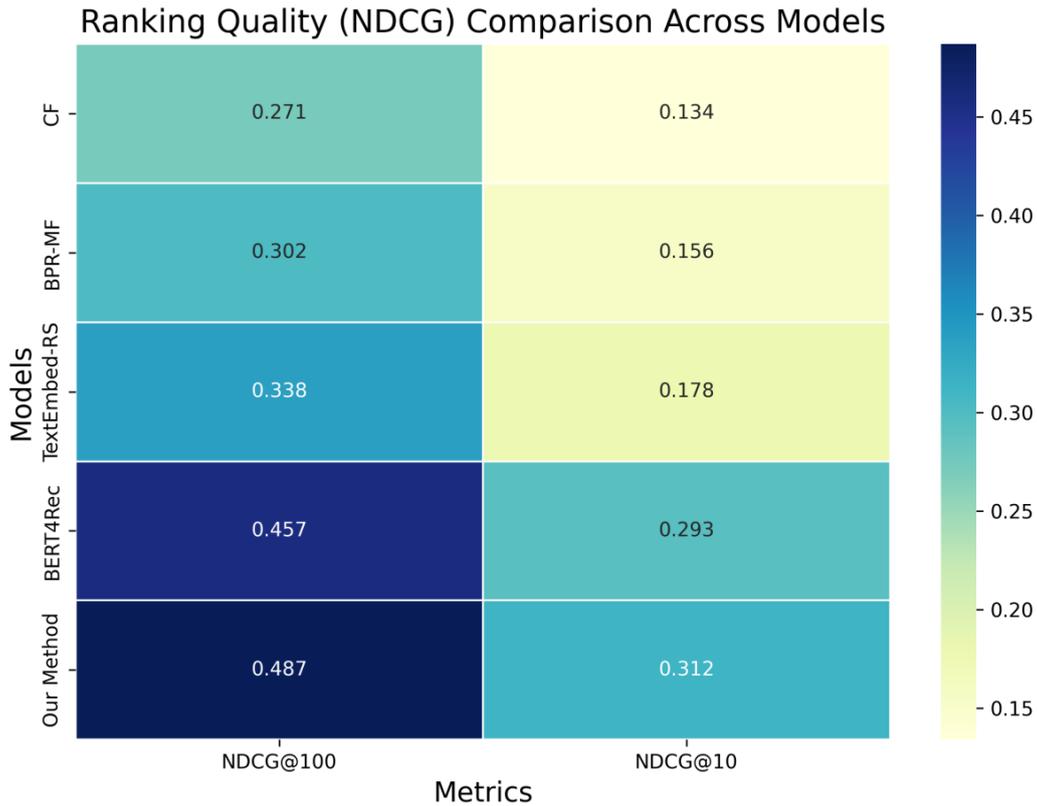

Figure 3: Ranking Quality (NDCG) Comparison Across Models

Another aspect of the system, which is serviceability, is presented in Table 1 as the average recommendation latency. Complexity of the computational operation when LLM fine-tuning and multi-source fusion are introduced is higher, however, the latency is still 182 ms on average, with a slight increase over BERT4Rec, i.e., 175 ms, thus meeting the requirement for e-commerce platforms to operate in real-time (under 200 ms).

Table 1: Average Recommendation Latency Across Models

| Model | Average Latency (ms) |
| --- | --- |
| CF | 34 |
| BPR-MF | 48 |
| TextEmbed-RS | 112 |
| BERT4Rec | 175 |
| Our Method | 182 |

Eventually, the introduced solution has shown significant improvement over currently existing systems in terms of different evaluation criteria. Every component contributes significantly to the system's performance, providing an efficient and practical approach to the e-commerce long-tail product recommendation problem [31].

## 4 DISCUSSION

### 4.1 Model Advantages, Disadvantages, and Applicability

Through our experiments and ablation analysis, we evaluated our long-tail–oriented large-scale language model in comparison with traditional recommendation approaches. While collaborative filtering offers simplicity and robustness for popular items, it performs poorly in cold-start and long-tail scenarios. Furthermore, Bo et al. highlighted that attention mechanisms struggle with sparse data and consequently lack the ability to effectively capture deep contextual semantics, which limits their adaptability in such recommendation environments [32].

The BPR-MF model provides intuitive ranking optimization and handles implicit feedback with moderate computational cost. However, it still suffers from data sparsity and fails to fully exploit textual features. Zhong et al., in their comparative analysis of custom and transfer learning models, similarly observed performance degradation in deep learning models under limited-data conditions, reinforcing the need for more data-efficient architectures [33].

Shallow text-embedding methods are inexpensive to implement and can capture basic semantic information, yet they lack contextual understanding and lead to limited recommendation diversity. Xu et al. emphasized, in their study on explainable AI in NLP, that insufficient context modeling hampers both generalization and interpretability, which is also a concern in recommendation systems relying solely on surface-level text features [34].

To address these challenges, our model leverages the deep semantic representation capabilities of a pre-trained language model, fuses collaborative and generative signals to balance precision and diversity, and employs a multi-stage ranking framework to optimize recall within acceptable latency. This design is particularly suited for large-scale e-commerce platforms where structured semantic features extracted from product descriptions and reviews can support real-time applications such as search ranking, homepage personalization, and marketing push. Zhu et al. proposed using diffusion priors to improve out-of-distribution detection in high-dimensional tasks, offering theoretical grounding for our approach in cold-start and generalization scenarios [35].

**4.2 Practical Application Prospects and System Scalability**

The rapid advancement of large-scale language models and cloud inference services makes LLM-based long-tail recommendations both practical and scalable for modern e-commerce. By encapsulating the LLM as a "text understanding and candidate generation" microservice—communicating via RPC or message queues—it integrates seamlessly into multi-stage pipelines while remaining decoupled and able to auto-scale under peak loads. As Su Pei-Chiang et al. (2022) demonstrated, their proposed quantum-inspired simplified swarm optimization algorithm for real-time task scheduling in multi-processor systems can effectively optimize system performance under high concurrency and peak loads, which is crucial for ensuring the auto-scaling capability of LLM microservices in such environments [36]. Companies can choose from trillion-parameter cloud models or lean distilled versions to balance recommendation quality and cost, offering flexibility for various applications. Yin Z et al. (2024), in their comparative study of CatBoost and XGBoost models for employee turnover prediction, provide valuable insights for selecting the best algorithm and optimizing recommendation systems, particularly in balancing recommendation quality and computational cost when dealing with dynamic data [37]. Beyond cold-start scenarios and new product launches, this approach supports seasonal promotions and personalized marketing by semantically aligning user intent with product attributes. Sun S et al. (2024), through their research on optimizing teaching methods with machine learning, provide a methodology that can be applied to personalize and optimize recommendation systems, especially in the context of seasonal promotions and new product launches [38]. Future extensions include multimodal fusion—ingesting images and videos—and online learning, where streaming data fine-tunes model weights to track evolving user interests, significantly enhancing user engagement. Duan Chenming et al. (2024), in their work on real-time psychological state prediction using BERT-XGBoost, provide a model for real-time user behavior and interest prediction, which can greatly improve personalization and real-time relevance in recommendation systems [39]. Continuous A/B testing and metric-driven tuning of fusion weights, candidate counts, and attention mechanisms ensure optimized performance and high adaptability. Zhang T et al. (2022), with their study on COVID-19 localization and recognition using Yolov5 and EfficientNet, demonstrate how multimodal data fusion can improve recommendation systems, particularly by integrating image and video content, thereby increasing recommendation accuracy and user satisfaction [40-42]. Leveraging multi-language and domain-specific LLMs further broadens applicability across global and vertical markets, driving user satisfaction and revenue growth. Sun S et al. (2024), by emphasizing the continuous optimization of system performance through machine learning and A/B testing, provide a framework for ensuring long-term improvements in recommendation systems, especially in global and vertical markets using multi-language and domain-specific LLMs [43].

**5 CONCLUSION**

This article demonstrates that leveraging LLMs as a source for recommendation models can effectively address many of the challenges associated with traditional recommendation methods in long-tail scenarios, thereby mitigating the problems of data scarcity and the cold-start effect. The proposed model has its approach of deep semantic retrieval to go from texts of the products to the vectors then combining these vectors with the user's behavior sequences. The so-called multi-source fusion recommendation model illustrates the method. The method has been validated by experiments, which demonstrate the method's superiority over the conventional recommendation systems in terms of recall, ranking quality, diversity, and long-tail coverage. Furthermore, it has been more effective in long tail products' exposure and hit rate. Despite the computational resources required, our method effectively enhances the recommendation performance for long-tail products, offering a significant improvement over traditional approaches.

In addition, the flexibility and scalability of the model allow it to adapt to various e-commerce domains and user types. By capturing subtle semantic nuances and integrating multi-dimensional user intent signals, our system contributes to the development of more intelligent, accurate, and diverse recommendation engines tailored for long-tail scenarios.